\documentstyle[aps,prd,epsf]{revtex}
\begin{document}
\draft
\title{Detecting supersymmetric Q-balls with neutron stars}
\author{Jes Madsen}
\address{Institute of Physics and Astronomy,\\
University of Aarhus, \\
DK-8000 \AA rhus C, Denmark\\
e-mail: jesm@dfi.aau.dk}
\date{16 May, 1998}
\maketitle

\begin{abstract}
Supersymmetric Q-balls trapped in neutron stars or white dwarfs 
may cause the stars to explode. Trapping of Q-balls
in neutron stars is shown to be less likely, but
trapping in neutron star progenitors more likely than hitherto assumed, 
making neutron stars very sensitive Q-ball ``detectors''.
White dwarfs only trap potentially dangerous Q-balls 
in a narrow parameter range.
\end{abstract}

\pacs{12.60.Jv, 97.60.Jd, 98.70.Rz, 98.80.Cq. Accepted for publication 
in PHYSICS LETTERS B}

Supersymmetric Q-balls \cite{coleman} with non-zero baryon number, $Q_B$, have
been suggested as a possible outcome of early Universe physics, and may
be closely related to the formation of net baryon number
\cite{kushap,enqvist}. Should they
survive, they would contribute to dark matter in the Universe today
\cite{kushap}.
Recently it was suggested that Q-balls trapped by neutron stars could
convert the star into supersymmetric phase, thereby reducing the stellar
mass until an explosion would set in as the minimal neutron star mass
was reached \cite{kstt,kkst}. Simulations using general relativistic
hydrodynamic codes indicate that such explosions would be observable as
mini-supernova events, releasing of order $10^{52}$ erg/s in
electron-antineutrinos and $10^{47}$ erg/s in keV $x$-rays
\cite{colal93,sumal98}. The events might also
be related to the mysterious
cosmic gamma-ray bursts \cite{kstt,kkst}. A similar Q-ball trapping scenario 
was suggested for white dwarf stars \cite{kstt,kkst}.

The discussion of astrophysical consequences of
supersymmetric Q-balls in many ways resemble
considerations related to other non-topological solitons carrying
baryonic charge, in particular lumps of strange quark matter, also known
as quark nuggets\cite{witten,dergla84}. 
The major difference in the context of astrophysical
constraints and applications stems from the fact that the mass of most
previously considered candidates grows proportionally to the baryon number,
whereas the mass of supersymmetric Q-balls grows only as $Q_B^{3/4}$.

Here several arguments developed for strange quark matter \cite{madsen88}
and other Q-balls \cite{madsen90}
will be generalized to the supersymmetric case. In particular it
will be shown that trapping of supersymmetric
Q-balls in the stellar precursors of
neutron stars is more likely than hitherto assumed, thereby
increasing the likelihood of neutron star conversion (in contrast,
trapping in the neutron star itself is shown to be significantly
suppressed relative to the expectations in \cite{kstt}). 
A single Q-ball present in the neutron star interior is
sufficient to ultimately transform the whole star. The time scale for
this transformation depends on the detailed physics, but is likely to be
small compared to the age of the Universe \cite{kstt}.
Thus stars could be very sensitive Q-ball ``detectors''.

Other limits previously derived for ``ordinary'' baryonic
Q-balls are not relevant for the supersymmetric counterparts. For
instance, electrically charged Q-balls present during the 
era of Big Bang nucleosynthesis may absorb neutrons and repel protons,
thereby reducing the important production of $^4$He
\cite{madris,madsen90}. This effect grows
in significance with the total surface area of Q-balls present, but since
the supersymmetric Q-balls are very compact relative to, e.g.\ strange
quark matter, a simple calculation shows, that no interesting limits on
supersymmetric Q-ball parameters can be derived in this manner.

Q-balls exist in the minimal supersymmetric extension of the standard
model as coherent states of squarks, sleptons and Higgs fields. In
theories with flat scalar potentials solitons with large baryon number
can be stable. For a potential $U(\phi )\sim m^4 ={\rm const}$ for large
scalar field $\phi$,
the mass of a supersymmetric Q-ball
is approximately given by \cite{kkst} 
$M_Q\approx 4\pi\sqrt{2} m Q_B^{3/4}/3 \approx 5924 {\rm
GeV} m_T Q_B^{3/4}$, with $m\approx\; $0.1--100 TeV, and $m_T\equiv m/1{\rm
TeV}$. A Q-ball is stable relative to nucleons of mass $m_n$ for $M_Q <Q_B m_n$,
giving the stability constraint $Q_B> 1.59\times 10^{15}m_T^4$. The
binding energy of a nucleon is $I_n\equiv m_n+M(Q_B)-M(Q_B+1)\approx
m_n[1-4729m_TQ_B^{-1/4}] \rightarrow m_n$ for $Q_B\rightarrow\infty$.

The radius of a Q-ball is $R_Q\approx Q_B^{1/4}/\sqrt{2}m\approx
1.40\times 10^{-17} m_T^{-1}Q_B^{1/4}$cm, giving a geometrical cross
section
\begin{equation}
\sigma_g =\pi R_Q^2\approx 6.11\times 10^{-34}m_T^{-2}Q_B^{1/2} {\rm cm}^2.
\label{eq:geo}
\end{equation}

Supersymmetric Electrically Charged Solitons (SECS)
with a net positive electrical charge in the interior
(neutralized by a surrounding cloud of electrons) will interact via
electrostatic scattering off matter. The cross section for this will be
the Bohr cross section
\begin{equation}
\sigma_B =\pi r_B^2\approx 10^{-16} {\rm cm}^2,
\label{eq:Bohr}
\end{equation}
unless $R_Q > r_B$, which happens for $Q_B> 2.68\times 10^{34}m_T^4$.
In that case, as well as for Supersymmetric Electrically Neutral Solitons
(SENS) the geometrical cross section, Eq.\ (\ref{eq:geo}) is the
relevant one.

Q-balls passing through matter (for example crossing through a star)
will be slowed down significantly after removing a total mass
comparable to their own. The mass removed by scattering is $M_S =\sigma D$,
where $D=\int \rho (x) dx$ is the total column density of the gas.
For Q-balls moving one stellar diameter through a star of mass $M_\star$ and
radius $R_\star$, the total column density is roughly
\begin{equation}
D\approx 5.0M_\star /R_\star^2 \approx 2\times 10^{12}M_\star
/M_{\odot}(R_{\odot}/R_\star )^2 {\rm g\; cm}^{-2} ,
\end{equation}
where subscript $\odot$ denotes solar values (the value for $D$ assumes
the star to be a polytrope of index 3, which is a reasonable
approximation for the applications below, and the Q-ball to be moving on
a radial trajectory, which is also a good approximation because of the
gravitational attraction from the star).

Thus stopping of a Q-ball requires $M_S >M_Q$. This corresponds to
stopping of Q-balls with baryon number below some critical $Q_S$ given
by
\begin{equation}
Q_S\approx 5.2\times 10^{21} m_T^{-4/3}
\left({\frac{M_\star}{M_\odot}}\right)^{4/3}
\left({\frac{R_\star}{R_\odot}}\right)^{-8/3} 
\end{equation}
for SECS scattering with the Bohr cross section, and 
\begin{equation}
Q_S\approx 1.8\times 10^{-4} m_T^{-12}
\left({\frac{M_\star}{M_\odot}}\right)^{4}
\left({\frac{R_\star}{R_\odot}}\right)^{-8} 
\label{eq:largesecs}
\end{equation}
for neutral Q-balls (SENS) and large SECS scattering with a geometrical
cross section.

For Q-balls in a monoenergetic distribution the accretion rate on
stars is
\begin{equation}
{\cal R}=2\pi GM_{\star}^2 n_\infty v_\infty^{-1}\frac{R_\star}{M_\star}
\left( 1+ {\frac{v_\infty^2R_\star}{2M_\star G}} \right) ,
\label{eq:totflux}
\end{equation}
where $v_\infty$ and $n_\infty\equiv\rho_\infty /M_Q$ are the
characteristic speed and number densities far from the accreting star.
Scaling to typical values for dark matter in the Galaxy we define
$v_\infty \equiv v_{250}\; 250\; {\rm km\; s}^{-1}$ and
$\rho_\infty \equiv \rho_{24}\; 10^{-24}\; {\rm g\; cm}^{-3}$ and get
\begin{equation}
{\cal R}=2.20\times 10^{26} {\rm s}^{-1} Q_B^{-3/4} m_T^{-1}
\rho_{24} v_{250}^{-1}
\frac{M_\star}{M_\odot}\frac{R_\star}{R_\odot} \left[ 1+0.164v_{250}^2
 \frac{M_\odot}{M_\star}\frac{R_\star}{R_\odot} \right] .
\label{eq:gravflux}
\end{equation}

Whereas geometrical accretion (second term in Eq.\ (\ref{eq:gravflux}))
dominates when Q-balls hit the Earth, leading to a flux of
\begin{equation}
F_{\rm Earth}=94  Q_B^{-3/4} m_T^{-1} \rho_{24} v_{250} 
 {\rm cm}^{-2} {\rm s}^{-1} {\rm sterad}^{-1},
\label{eq:earthflux}
\end{equation}
accretion on stars is mainly
governed by gravitational capture (first term in Eq.\
(\ref{eq:gravflux}); the gravitational analog of the Rutherford cross
section. In contrast, Ref.\ \cite{kkst} apparently 
assumes geometrical accretion on stars and underestimates the rate of
SENS accretion on neutron stars by a factor $10^4$). 
The total number of Q-balls hitting a star in time $t$ is therefore
\begin{equation}
N\approx 6.9\times 10^{33} \frac{t}{1 {\rm yr}} \rho_{24} v_{250}^{-1} m_T^{-1}
Q_B^{-3/4} \frac{M_\star}{M_\odot}\frac{R_\star}{R_\odot} ,
\end{equation}
the total baryon number $Q_{\rm accrete}=NQ_B$, and the total mass 
\begin{equation}
M_{\rm accrete}\approx 3.7\times 10^{-20} M_\odot \frac{t}{1 {\rm yr}} 
\rho_{24} v_{250}^{-1} \frac{M_\star}{M_\odot}\frac{R_\star}{R_\odot} .
\end{equation}

A star is hit by at least one Q-ball in time $t$ if Q-balls with
$Q_B\leq Q_1$ contribute with density $\rho_{24}$, where
\begin{equation}
Q_1 = 1.3\times 10^{45}\left({t}\over {1 {\rm yr}}\right)^{4/3}
\rho_{24}^{4/3}v_{250}^{-4/3} m_T^{-4/3}\left( M_\star \over M_\odot
\right)^{4/3}\left( R_\star \over R_\odot \right)^{4/3} .
\label{eq:Q1}
\end{equation}

The equations above are crude approximations in several respects. The
true distribution of Q-balls would not be monoenergetic. Velocity peaks in 
the spectrum would appear as a result of the gravitational relaxation
processes during galaxy formation \cite{ipssik92,sikal95}, and some
enhancement (by factors of a few \cite{fretur83}) 
in the local flux hitting the Earth will result because of
Q-balls gravitationally captured in the Solar System \cite{fretur83,dimal82}. 
While such effects are very important for detailed studies of the dark
matter distribution in our Galaxy, they are also very model dependent
and sufficiently small that the simple average flux equations used above
suffice for the present purpose.

Ref.\ \cite{madsen88} considered the 
trapping of lumps of strange quark matter in
neutron stars and especially their stellar progenitors, showing that
very strong limits could be placed on their cosmic abundance under the
assumption that some pulsars had to be ``ordinary'' neutron stars rather
than converted into quark stars 
(or conversely, that if strange quark matter is stable,
then all ``neutron'' stars are actually quark stars, since some quark
lump pollution of interstellar space is unavoidable, for instance from
quark star collisions). In Ref.\ \cite{madsen90}
this argument was generalized to other baryonic Q-balls with mass
proportional to baryon number.

Kusenko {\it et al.\/} \cite{kstt,kkst} consider the even more spectacular
possibility that supersymmetric Q-balls being
trapped by neutron stars may eat up the neutrons, thereby reducing the
neutron star mass
below the stability limit of roughly $0.1 M_\odot$, leading to an explosion
that might be related to gamma-burst events.

Whereas capture in neutron stars is possible in certain ranges of Q-ball
parameters, it will be shown below that several effects
make trapping in neutron stars less likely than envisaged in \cite{kstt}. In
contrast, trapping in neutron star progenitors will be shown to be much
more likely than estimated by Kusenko {\it et al.\/}, just like the
cases studied in \cite{madsen88,madsen90}.
The corresponding trapping of electrically neutral Q-balls (SENS) in
white dwarf stars is demonstrated not to take place for most parameters, 
contrary to the findings in \cite{kstt}. (Only neutral Q-balls are
potentially dangerous to white dwarf survival, as white dwarfs contrary
to neutron stars do not possess free neutrons capable of interacting with SECS).

Two aspects were left out of the discussion of Q-ball trapping in
neutron stars in \cite{kstt}. First the fact that Q-balls are extremely
relativistic when hitting the star, and second the fact that the crust
of a neutron star consists of an ion lattice with a substantial
structural resistance, capable of preventing light Q-balls from reaching
the central regions of the star.

Concerning the first point, gravitational acceleration causes a typical
Q-ball to reach the stellar surface with a velocity $v\approx (2GM_\star
/R_\star )^{1/2}$ and a kinetic energy per baryon of 1TeV $m_T Q_B^{-1/4}$
for typical neutron star parameters. This means that Q-balls near the
stability limit will experience a highly inelastic collision where the
Q-ball is heated and perhaps (partly) disintegrates. Since the binding
energy per baryon increases with $Q_B$, this disintegration is only
important for $Q_B$ close to the lower limit for stability.

The second point, Q-balls (or fragments from the disintegration just
discussed) being trapped in the crustal lattice of the
neutron star, thus being unable to reach the central regions and convert
the star into supersymmetric phase, is more important, 
since the star is only molten
in the first few months following formation in a supernova explosion.
The column density of the $5\times 10^{28} {\rm g}$ solid crust above
the neutron drip region, below which free neutrons are available for absorption
in Q-balls, is $D_{\rm crust}\approx 4\times 10^{15} {\rm g\; cm}^{-2}$.
Only charged Q-balls with $Q_B > 1.3\times 10^{26}m_T^{-4/3}$ and
neutral Q-balls with $Q_B > 2.9\times 10^9 m_T^{-12}$ penetrate the
crust without stopping. Q-balls that are stopped may sink to the interior
if gravity $GM_\star M_Q/R_\star^2$ exceeds lattice resistance
$\epsilon\sigma_Q$, where the structural energy density for crystalline
material in neutron star crusts consisting of ions with mass and charge
$A$, $Z$, and density $\rho$ is roughly $\epsilon\approx 1\times 10^{13}
{\rm ergs\; cm}^{-3} \rho^{4/3} Z^2A^{-4/3}$. For an iron lattice at
$10^{11} {\rm g\; cm}^{-3}$, $\epsilon_{28}\equiv \epsilon /10^{28}
{\rm ergs\; cm}^{-3}\approx 1$.

Stable neutral Q-balls will ultimately sink to the center because of
gravity (unless $Q_B<4\times 10^2 m_T^{-12}\epsilon_{28}^{4} (R_\star
/10 {\rm km})^8 (M_\odot /M_\star )^{4}$),
but charged Q-balls are trapped in the crust if $Q_B< 6.4\times
10^{23}m_T^{-4/3}\epsilon_{28}^{4/3} (R_\star /10 {\rm km})^{8/3}
(M_\odot /M_\star )^{4/3}$ and will not pose a threat to the survival
of the neutron star.

Contrary to the assumption in \cite{kstt} the most important trapping mechanism
for charged supersymmetric Q-balls over large ranges of $Q_B$ is
actually trapping in the supernova progenitor, so that Q-balls are
present at the time of neutron star formation. Figure 1 shows the
flux of Q-balls
reaching the Earth according to Eq.\ (\ref{eq:earthflux}) as a function
of baryon number for those contributions to galactic
halo density where at least one Q-ball is trapped in different
phases of stellar evolution (just one Q-ball is needed to start the
conversion of the neutron star into supersymmetric phase). 
The curve marked MS corresponds to Q-ball trapping in a $10 M_\odot$
main sequence star of radius $6.3 R_\odot$, with a main-sequence
lifetime of $5\times 10^7 {\rm years}$, 
PMS marks trapping in the post-main-sequence phase
just prior to a supernova explosion (estimated as 2 months capture in
the $1 M_\odot$ stellar core of radius $10^{-2} R_\odot$), 
and NS denotes trapping in $10^9$
years in a $1.4 M_\odot$, solid crust neutron star of radius 10 km. 
The horizontal dashed line would be
the corresponding sensitivity to capture during the first few months
after the supernova, where the neutron star is molten, and light Q-balls
are not trapped in the crust. There will be some lower cut-off in 
$Q_B$ where Q-balls
are destroyed because of the relativistic, inelastic collision with the
neutron star, but this limit has not been calculated in detail. 
Instead, the vertical line in the left part of the figure is the
ultimate cut-off, corresponding to the lower limit of Q-ball stability
for $m_T=0.1$. Also marked along the first axis are the lower cut-offs
for $m_T=1$, 10, and 100. The first axis has been scaled to
$Q_Bm_T^{4/3}$ so that all flux curves are independent of $m_T$, except
the stability cut-off, and the experimental results (EXP), which happen
to be valid for baryon numbers down very close to the stability cut-off
for a given choice of $m_T$. The experimental flux-limits are derived
from the discussion in Ref.\ \cite{baksan} on the magnetic monopole
searches with the Baksan detector, and should be valid for SENS as well
as SECS, whereas the other limits given in Figure 1 are valid for SECS
only. These experimental limits may be improved by some 3 orders of
magnitude if other monopole search results are properly reinterpreted to
be valid for Q-balls, but the
astrophysical (stellar) ``detectors'' are seen to be potentially
much more sensitive because of the large surface areas and long
integration times.

For comparison Figure 1 also shows the expected flux if Q-balls of a
given $Q_Bm_T^{4/3}$ were responsible for the dark matter in our Galaxy
($\rho_{24} v_{250}=1$ in Eq.\ (\ref{eq:earthflux})).

Because of the small geometrical cross section of neutral supersymmetric
Q-balls, the methods discussed above are much less sensitive to SENS
(c.f.\ Fig.\ 2).
In fact, SENS pass through main sequence stars without being
significantly affected, and only neutron stars are able to stop neutral
Q-balls (with $Q_B <4\times 10^{35}m_T^{-12}$), except for $m_T<1$,
where it may also happen in white dwarfs and in post-main-sequence stars.
Only neutral Q-balls with $Q_B<10^{12} m_T^{-12}$ are stopped in these
systems, so white dwarfs are only endangered by trapped SENS if $m_T$ is
small (Q-balls with $Q_B<1.59\times 10^{15}m_T^4$ are unstable). This 
differs from the results of Ref.\ \cite{kstt}.

Trapping of supersymmetric Q-balls in different stellar types has been
studied and related to the suggestion that neutron stars or white dwarfs
may explode when transformed into a scalar condensate by the accumulated
Q-balls. While accumulation of electrically neutral Q-balls 
in white dwarfs is important only for a very narrow range of parameters,
capture of charged and neutral Q-balls in neutron star progenitors
complements the capture in the neutron stars themselves to make neutron
stars potentially very efficient Q-ball detectors---many orders of
magnitude more flux-sensitive than groundbased experiments. A major
uncertainty in the conclusion is the actual lifetime of a neutron star
being transformed by the Q-balls. Only preliminary estimates have been
given \cite{kstt}, showing that it may be in a range of relevance for
cosmic gamma-ray bursts. Another unknown (which would work to increase
the effective ``capture rate'') is the amount of Q-balls (if any)
present in the cloud forming the neutron star progenitors. 
This depends completely on the prior evolution history of the Galaxy
and the interstellar gas. But regardless of the uncertainties, it
appears that major windows of supersymmetric Q-ball parameters do
provide interesting consequences for neutron star evolution.

\acknowledgments
This work was supported in part by the Theoretical Astrophysics Center
under the Danish National Research Foundation.

\begin{figure}
\epsfxsize=12truecm\epsfbox{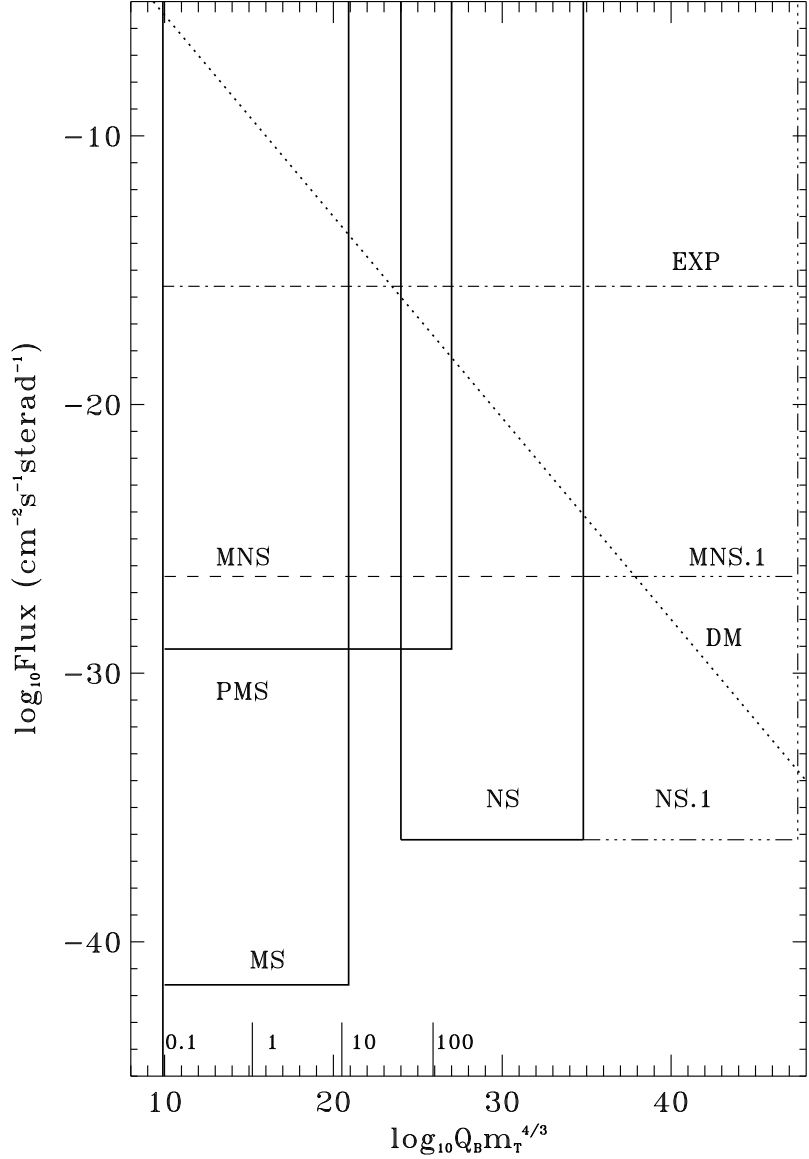}
\caption{Flux of SECS Q-balls hitting the Earth versus Q-ball baryon number
(multiplied by $m_T^{4/3}$ to make most curves independent of $m_T$),
for which at least one charged Q-ball is trapped in upper
main sequence stars (MS), post-main-sequence stars (PMS), 
neutron stars (NS), and molten neutron stars (MNS). For
comparison limits from detector searches are denoted by EXP.
The diagonal curve (DM) is the upper flux limit
corresponding to galactic dark matter. For $m_T=0.1$ SECS capture
in neutron stars is governed by Eq.\ (\protect{\ref{eq:largesecs}}); the curves
NS.1 and MNS.1. See text for further explanations.
}
\label{fig1}
\end{figure}
\begin{figure}
\epsfxsize=12truecm\epsfbox{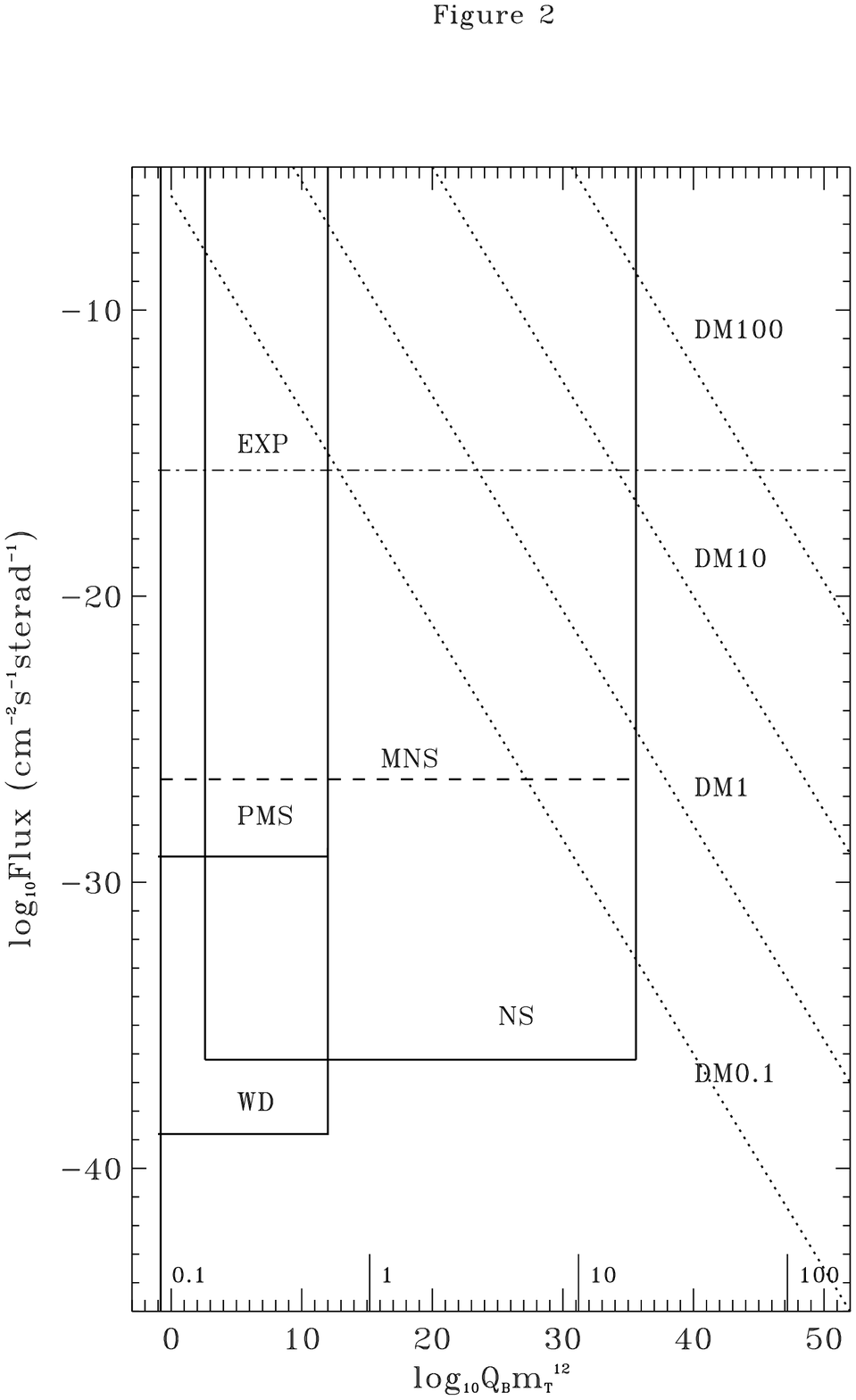}
\caption{Flux of SENS Q-balls hitting the Earth versus Q-ball baryon number
(multiplied by $m_T^{12}$ to make most curves independent of $m_T$).
Most of the notation as in Fig.\ 1. The numbers after DM for the dark matter
curves indicate the value of $m_T$. WD-curve shows trapping
in a $10^9$ year old white dwarf of mass $1 M_\odot$, radius $10^{-2}
R_\odot$. See text for further explanations.
}
\label{fig2}
\end{figure}

\end{document}